# Interference between two independent electrons: observation of two-particle Aharonov-Bohm interference

I. Neder[1], N. Ofek[1], Y. Chung[2], M. Heiblum[1], D. Mahalu[1], and V. Umansky[1]

[1]*Braun Center for Submicron Research, Department of Condensed Matter Physics, Weizmann Institute of Science, Rehovot 76100, Israel*

[2]*Department of Physics, Pusan National University, Busan 609-735, Republic of Korea*

**Very much like the ubiquitous quantum interference of a single particle with itself, quantum interference of two independent, but indistinguishable, particles is also possible. This interference is a direct result of quantum exchange statistics, however, it is observed only in the joint probability to find the particles in two separated detectors. Here we report the first observation of such interference fringes between two independent and non-interacting electrons in an interferometer proposed by Yurke *et al.* [1] and Samuelsson *et al.* [2]. Our experiment resembles the 'Hanbury Brown and Twiss' (HBT) experiment, which was performed with classical waves [3,4]. In the experiment, two independent and mutually incoherent electron beams were each partitioned into two trajectories. The combined four trajectories enclosed an Aharonov-Bohm (AB) flux (but not the two trajectories of a single electron (Fig. 1a)). While individual currents were found to be independent of the AB flux, as expected, the cross-correlation between current fluctuations in two opposite points across the device exhibited strong AB oscillations. This is a direct signature of orbital entanglement between two electrons even though they never interact with each other.**

We report the observation of 'second order interference' of electrons, namely, where currents of two independent electron streams - at two different locations - exhibit phase dependent correlation. This is instead of the familiar interference between the amplitudes of wave functions. In optics, attempts to perform similar experiments with single photons encountered fundamental difficulties such as: generating indistinguishable photons in a narrow band of frequencies and synchronizing their arrival in time. Hence, experiments concentrated thus far on proving bunching of single photons (bosons) via coincidence measurements (the Hong-Ou-Mandel interference effect) [5 and references in it]. Similar experiments with electrons are rather scarce. The experiments that claimed to be HBT like merely reported the negative correlation between transmitted and reflected electron currents from an electronic beam splitter [6-9]; very much like measurements of quantum shot noise relying on anti bunching of fermions. While one can find earlier proposals [10,11] for the realization of orbital entanglement of electrons using edge channels in the quantum Hall effect regime [12], a practical interference-type electronic HBT-like experiment was proposed recently by Samuelsson *et al.* [2]. It was based on the newly developed electronic Mach-Zehnder interferometer in a two-dimensional-electron-gas (2DEG) [13]; utilizing edge channels. In many ways an experiment with electrons is easier than with photons. Injecting electrons from an extremely cold and degenerate fermionic reservoir produces a highly ordered beam of electrons that is totally noiseless [14]; hence, a high coincidence rate is achieved without a need for synchronizing electrons' arrival times. Moreover, since the coherence length of the electrons ('wave packet width' or 'spatial size') is determined by the applied voltage (at nearly zero temperature), a very small applied voltage assures a single electron at a time in the interferometer, preventing electron-electron interaction. Moreover, since each electron has definite energy (Fermi energy) and momentum (Fermi momentum), the electrons are indistinguishable. However, the applied small voltage leads to exceedingly small electrical current and to minute fluctuations; making the measurements extremely difficult to perform.

The schematic of our experiment is shown in Fig. 1(a) [1]. Two independent, separated, sources of electrons (S1 & S2) inject ordered, hence noiseless, electrons toward each other. Each stream passes through a beam splitter (A & B), and splits into two negatively correlated partitioned streams (if an electron turns right a hole is injected to the left). Both sets of the two partitioned streams join each other at two additional beam splitters (C & D), interfere there and generate altogether four streams that are collected by drains D1 through D4. Hence, every electron emitted by either S1 or S2 eventually arrived at one of the four drains. Consider now the event where one electron arrives at D2 and the other arrives at D4. There are two quantum mechanical probability amplitudes contributing to this event: S1 to D2 and S2 to D4; or, alternatively, S1 to D4 and S2 to D2. These two "two-particle" events can interfere because they are indistinguishable. Since in the two possible events the electrons travel along different paths - picking thus different phases - the joint probability for one to arrive at D2 and the other at D4 contains the total phase of all paths - as we show below.

The two wave functions, corresponding to the *incoming* states from each of the two sources $\Psi_{Si}$, can be expressed in the basis of the *outgoing* states at the four drains $\psi_{Dj}$. Assuming, as in the experiment, that every beam splitter is half reflecting and half transmitting, its unitary scattering matrix $M$ (that ties the input and output states) can be taken as: $M = \begin{bmatrix} r & t \\ t' & r' \end{bmatrix} = \frac{1}{\sqrt{2}} \begin{bmatrix} i & 1 \\ 1 & i \end{bmatrix}$. Considering the phases of the four possible paths $\phi_1,..,\phi_4$:

$$\Psi_{S1}(x) = \frac{1}{2}\left[ie^{i\phi_1}\psi_{D1}(x) - e^{i\phi_1}\psi_{D2}(x) + ie^{i\phi_2}\psi_{D3}(x) + e^{i\phi_2}\psi_{D4}(x)\right] , \quad (1a)$$

$$\Psi_{S2}(x) = \frac{1}{2}\left[ie^{i\phi_3}\psi_{D1}(x) + e^{i\phi_3}\psi_{D2}(x) + ie^{i\phi_4}\psi_{D3}(x) - e^{i\phi_4}\psi_{D4}(x)\right] . \quad (1b)$$

Since each electron is not allowed to interfere with itself only particle statistics could cause interference. Because of the fermionic property of electrons, the total two-particle wave function must be the anti-symmetric product of Eq. 1a and Eq. 1b:

$$\Psi_{total}(x_1,x_2) = \frac{1}{\sqrt{2}}\left[\Psi_{S1}(x_1)\Psi_{S2}(x_2) - \Psi_{S2}(x_1)\Psi_{S1}(x_2)\right], \quad (2)$$

with $x_1$ and $x_2$ any two locations in the interferometer. Substituting Eq. 1 in Eq. 2 leads to 24 terms, expressing the probability amplitude for one electron at $x_1$ and another at $x_2$. Since we wish to concentrate on correlations between drains we write $\Psi_{total}$ using the notation $\psi_{DiDj} \equiv \frac{1}{\sqrt{2}}\left[\psi_{Di}(x_1)\psi_{Dj}(x_2) - \psi_{Dj}(x_1)\psi_{Di}(x_2)\right]$ for an anti-symmetric state, in which one electron heads to $Di$ and another to $Dj$. The two-particle wavefunction is:

$$\Psi_{total}(x1,x2) = \frac{i}{2}\left(e^{i(\phi_1+\phi_3)}\psi_{D1D2} - e^{i(\phi_2+\phi_4)}\psi_{D3D4}\right) + $$
$$+ \frac{i}{2}e^{\frac{i}{2}(\phi_1+\phi_2+\phi_3+\phi_4)}\left[\sin\left(\frac{\Phi_{total}}{2}\right)(\psi_{D2D4} - \psi_{D1D3}) - \cos\left(\frac{\Phi_{total}}{2}\right)(\psi_{D2D3} + \psi_{D1D4})\right] , \quad (3)$$

with $\Phi_{total} = \phi_1 - \phi_3 + \phi_4 - \phi_2$, which is exactly the total accumulated phase going anti-clockwise along the four trajectories of the two particles.

Equation 3 describes the *two-particle interference effect*, with the absolute value squared of the prefactor of $\psi_{DiDj}$ the joint probability to have one electron at *Di* and one at *Dj*. Concentrating on the correlation between D2 and D4, one can deduce from Eq. 3 the following: (a) Two electrons never arrive at the same drain (Pauli exclusion principle); (b) The first part suggests that there is a 50% chance for two electrons to arrive at the same 'side' simultaneously, namely, at D1 & D2, or at D3 & D4, but never at D2 & D4. (c) The second part suggests that there is a 50% chance for two electrons to arrive at opposite 'sides', namely, one at D1 or D2 and the other at D3 or D4; however, the exact correlation depends on $\Phi_{total}$. When $\Phi_{total}=\pi$, $\sin^2(\Phi_{total}/2)=1$ and two electrons arrive at (D1, D3) or at (D2, D4), but when $\Phi_{total}=0$, $\cos^2(\Phi_{total}/2)=1$ and the complementary events take place. (d) Combining all events in the two parts of the total wavefunction, one finds for $\Phi_{total}=0$ a perfect anti-correlation between the arrival of electrons in D2 and in D4; however, for $\Phi_{total}=\pi$ there is 50% chance of anti-correlation (first part) and 50% chance of positive correlation (second part) - hence, zero correlation! The time-averaged cross-correlated signal of the current fluctuations in the two drains is proportional to the probability of the correlated arrival of electrons in these drains. Varying the total phase should result in a negative oscillating CC signal between current fluctuations in D2 and in D4. The quantitative estimate of the amplitude of that CC signal is discussed later.

Figure 1(b) describes the realization of the experiment. The two-particle interferometer is shown split in the center, resulting in an upper and lower segments; each is a simple optical Mach-Zehnder interferometer (MZI) [15]. An electronic version of the MZI had been recently fabricated and studied [13,16,17]. A quantizing magnetic field (~6.4T) brings the 2DEG into the QHE state, and at filling factor one. The current is carried by a single edge channel along the boundary of the sample [12]. Being a chiral one-dimensional object, the channel is highly immuned against back scattering and dephasing. The layout of the two-particle interferometer is described in Fig. 1(c), with the SEM micrograph of the actual device shown in Fig. 1(d). The two MZIs can be separated from each other with a 'middle gate' (MDG). When it is closed, each MZI can be tested independently for its coherence and the AB periodicity. A quantum point contact (QPC), formed by metallic split gates, functions as a beam splitter while ohmic contacts serve as sources and drains. In this configuration the phase that is accumulated along the four trajectories is the AB phase, namely, $\Phi_{total}=\varphi_{AB}=2\pi BA/\Phi_0$, with $B$ the magnetic field and $A$ the area enclosed by the four paths ($\Phi_0=4.14\times10^{-15}$ Tm$^2$ is the flux quantum) [18] - hence, the title of this work. Look for example at the upper MZI of the separated two-particle interferometer (Fig. 1(b)). An edge channel, emanating from ohmic contact S1, is split by QPC1 to two paths that enclose a high magnetic flux and join again at QPC2. The phase dependent transmission coefficient from S1 to D2 is:

$$T_{MZI} = \left| t_{QPC1}t_{QPC2} + e^{i\varphi_{AB}} r_{QPC1}r_{QPC2} \right|^2 = T_0 + T_\varphi \cos\varphi_{AB} \quad , \qquad (4)$$

where $t$ and $r$ are the transmission and reflection amplitudes of the QPCs and the visibility is defined as $\nu=T_\varphi/T_0$. The AB phase was controlled by the magnetic field and the 'modulation gate' (MG1 or MG2) voltage $V_{MG}$, which affected the enclosed area by the two paths.

Figure 2 displays the measured conductance of the two separated MZIs (defined as $i_D/V_S=T_{MZI}(e^2/h)$, where $i_D$ the AC current in the drain and $V_S$ the applied AC voltage at the source, with $e^2/h$ the edge channel conductance). Pinching off MDG, the QPCs were tuned to transmission 0.5 and the AC signal was measured at D2 and D4 as function of the modulation gate voltage $V_{MG}$ and magnetic field. As $V_{MG}$ was scanned repeatedly the magnetic field decayed unavoidably (since the superconducting magnet is not ideal), at a rate of ~1.4G/hour. Hence, the interference pattern was 'tilted' in the two-coordinate plane of $V_{MG}$ and time (magnetic field), with two basic AB periods for each MZI [13]. Apparently, the seemingly identical MZIs had different periodicities: 1 mV & 80 min in the upper MZI and 1.37 mV & 87 min in the lower MZI (the asymmetry resulted from misaligning the QPCs and MGs). In the two MZIs we found visibilities 75-90%; by far the highest ever measured in an electron interferometer. The high visibility was likely to result from the smaller size of the MZIs [12,16,17]; hence, dephasing mechanisms such as flux fluctuations or temperature smearing were less effective. Moreover, the high quality 2DEG assured a better formation of 1d-like edge channels and better overlap of particle wave functions.

We then discharged the 'middle gate', thus opening it fully and turning the two MZIs to a single two-particle interferometer. The conductance at D2 and D4 were now found to be independent of the AB flux, with a visibility smaller than the background (<0.1%). This is expected as each electron did not enclose anymore an AB flux.

We turn now to discuss the current fluctuations, namely, the shot noise in D2 and in D4. Feeding a DC current in S1, the low frequency spectral density of the shot noise in the partitioned current (by QPC1) at D2 and at D4 (with QPC2 closed and QPC3 and MDG open) was measured. Its expected value (up to some temperature corrections) is $S_{D2}=2eI_{S1}T_{QPC1}(1-T_{QPC1})=0.5eI_{S1}$ (A$^2$/Hz) for $T_{QPC1}=|t_{QPC1}|^2=0.5$ [14]. The current fluctuations in the drain were filtered by a LC circuit, with 60kHz bandwidth around a center frequency ~0.8 MHz, and then amplified by the cold amplifier, followed by a room temperature amplifier and a spectrum analyzer. In order to calibrate the CC measurement, we performed three noise measurements: (a) noise measured at D2; (b) noise measured at D4; (c) noise measured by cross-correlating the current fluctuations at D2 and at D4 (by an analogue home-made correlating circuit). Measurements (a) and (b) both led accurately to the expected result above (they are anti-correlated and equal signals), which were used to calibrate measurement (c). An electron temperature of ~10mK was deduced from these measurements [19].

We were ready at this point to measure the two-particle CC. All four QPCs were tuned to $T_{QPC}=0.5$ while the 'middle gate' was left open, hence, turning the two MZIs to a single two-particle interferometer. Equal DC voltages were applied to sources S1 and S2 with two separated power supplies $V_{S1}=V_{S2}=7.8\mu V$ ($I_{S1}=I_{S2}=0.3nA$). For that voltage there is at most a single electron in each of the four trajectories of the interferometer (the wave packet's width, 15-30μm, estimated from the current and the estimated drift velocity ~3-6×10$^6$ cm/sec, is bigger then the interferometer's path length, being ~8μm). This guaranteed a stronger overlap between the wave functions of the two electrons and minimized Coulomb interaction among the electrons (thus eliminating non-linear effects in the interferometer [16]). The measured fluctuations in D2 and D4 were averaged over some 30,000 electrons, amplified by two separate amplification channels (each fed by its own power supply), and finally cross-correlated. In order to verify flux insensitivity in each drain separately, we first measured the shot noise in D2 and in D4 as function of the magnetic flux (varying MG voltage and magnetic field). The noise, with a spectral density of $S=0.5eI_S \cong 2.4 \times 10^{-29}$ A$^2$/Hz, was found to be featureless. For further assurance, a 2D fast Fourier transform (FFT) of the measurements was calculated with the results shown in Figs. 3(a) & 3(b). Again, the transforms were featureless and without any feature above our measurement resolution of ~2×10$^{-31}$ A$^2$/Hz; confirming the absence of flux periodicity in the noise (as was found also in the transmission).

We estimate now the expected magnitude of the CC signal from Eq. 3. When $I_{S1}=I_{S2}=I$ and $\Phi_{total}=0$, a maximum anti-correlation signal $S_{D2D4}=\langle \Delta I_{D2} \cdot \Delta I_{D4} \rangle$ is expected. It can be shown that the expected value, for a 100% visibility, is the same as that of the noise of a single QPC, that is $S_{QPC}=2eIT_{QPC}(1-T_{QPC})$, or $0.5eI$ (for $T_{QPC}=0.5$). Since for $\Phi_{total}=\pi$ the CC signal is expected to vanish, we may conclude that the CC signal should oscillate with $\Phi_{total}$, $S_{D2D4} = -0.25eI(1-\sin\Phi_{total})$, with amplitude 1.2×10$^{-29}$ A$^2$/Hz for $I$=0.3nA.

Without currents in the sources, the CC signal was featureless (the background), with an average over the 2D FFT $\sim 2\times 10^{-31}$ $A^2$/Hz (not shown). The CC measurement with $I$=0.3nA is shown in Fig. 4. Already in the raw data in Fig. 4(a) the AB oscillations are visible. In the 2D FFT in Fig. 4(b) one sees a sharp peak corresponding to a period of 0.58 mV in $V_{MG}$ (with the same voltage applied to MG1 and MG2) and a 42.5 min in time (proportional to the magnetic field decay). The square root of the integrated power under the FFT peak (the amplitude of the AB oscillations) is $3.0\times 10^{-30}$ $A^2$/Hz. Moreover, we could directly resolve the AB oscillations as function of $V_{MG}$ and time separately by coherent time averaging. Since the magnetic field decayed in time, thus adding continuously an AB phase, this extra phase could be compensated for by shifting subsequent scans in $V_{MG}$ according to the decay rate found in the 2D FFT; leading to the negative oscillatory CC fringes shown in the left panel of Fig. 4(a). Similarly, the oscillations as function of magnetic field had been extracted (right panel, Fig. 4(a)). In Fig. 4(c) we provided the vector representation of the periodicities (inverse of periods) of each individual MZI (from Fig. 2) and that of the two-particle interferometer; the last being, quite accurately, the sum of the two! This is expected, as the rate of change of the AB flux of the two-particle interferometer is the sum of the rates of the two MZIs.

Compared with the expected value of the CC oscillations amplitude, $1.2\times 10^{-29}$ $A^2$/Hz, we measured an amplitude of $3.0\times 10^{-30}$ $A^2$/Hz. The results are quite accurate since the measurements had been repeated a few times and over long periods of integration times; lowering the uncertainty to below $10^{-31}$ $A^2$/Hz. A few factors could lead to the lower CC signal: (a) Although we have no theory, it is likely that the lower visibility in each MZI will lower the CC signal by $v_{MZI1} \times v_{MZ2}$. While the visibilities at zero applied DC voltage were ~80% (see Fig. 2), the visibilities at the applied DC voltage $V_S$=7.8μV were found to be ~70% [16]; (b) Our finite temperature (~10mK) will lower the shot noise by ~22%, affecting similarly the CC signal. These two effects alone will lower the expected CC signal to ~$4.6 \times 10^{-30}$A$^2$/Hz - about 1.5 times higher than the measured one. This discrepancy it still not understood.

Our direct observation of interference between independent particles provides a reliable scheme to entangle separate, but indistinguishable, quantum particles. This demonstration, done with electrons, reproduces the original Hanbury Brown and Twiss experiments [3,4], which were performed with classical waves. It is of a fundamental value - being at the core of multiple particles wavefunction. Our scheme has the potential to test Bell inequalities [1,2]; however, taking into account the finite temperature, it seems that the possibility to violate Bell inequalities in our measurements (with a visibility of merely 25%) requires further theoretical analysis.

**Acknowledgement**

The work was partly supported by the Israeli Science Foundation (ISF), the Minerva foundation, the German Israeli Foundation (GIF), the German Israeli Project cooperation (DIP), and the Ministry of Science - Korea Program. YC was supported by KRISS and KICOS and NCoE at Hanyang University through a grant provided by the Korean Ministry of Science & Technology and by PRCP funded by KRF.

**Figure Captions:**

**Fig. 1.** The two-particle Aharonov-Bohm interferometer. (a) Schematics of the interferometer. Sources S1 and S2 inject particles, which split by beam splitters A & B, later to recombine at beam splitters C & D. Each particle can arrive at any of four different drains D1 through D4. Each of the four trajectories accumulates phase $\phi_i$. (b) By breaking the interferometer in the center, two separate Mach-Zehnder interferometers (MZI) are being formed. The MZIs are the building block of the two-particle interferometer. (c) A detailed drawing of the interferometer. It was fabricated on a high mobility GaAs-AlGaAs heterostructure, with a 2DEG buried some 70 nm below the surface (carrier density $2.2 \times 10^{11}$ cm$^{-2}$ and low temperature mobility $5 \times 10^6$ cm$^2$/v-sec). Samples were cooled to ~10mK electron temperature. Quantum point contacts served as beam splitters and ohmic contacts as sources and drains. Tuning gates MG1 & MG2 changed the area and thus the flux in the interferometer, and gate MDG separated the interferometer into two MZIs. Metallic air bridges connected drains D1 & D3 to the outside, where they were grounded. Currents at D2 & D4 were filtered first by a LC circuit (tuned to 0.8 MHz and 60 kHz bandwidth) and then amplified by a cold preamplifier (at 4.2 K). (d) SEM micrograph of the actual sample. Air bridges were used to contact the small ohmic contacts, the split gates of the QPCs, and the 'middle gate' MDG.

**Fig. 2.** Color plot of the conductance of the two separate MZIs as function of the MG voltage and magnetic field that decayed in time. Strong AB oscillations

dominate the conductance with visibilities of ~80% each. A 2D FFT in the inset provides the periodicity in MG voltage and in time ($V_{MG}$, Time).

**Fig. 3.** Analysis and 2D FFT of auto-correlation (shot noise) for an open 'middle gate'. (a) & (b) 2D FFT of shot noise measurements in D2 & in D4. The noise is totally featureless with no sign of AB oscillations above the background.

**Fig. 4.** Cross-correlation of the current fluctuations in D2 and D4. (a) Main: 2D color plots of the raw data as function of the MG voltage and time (magnetic filed). The periodicity is already visible in the raw data. Right panel: Coherent averaging of some 50 traces as function of $V_{MG}$, by correcting for the added phase due to the decaying magnetic field (see text). Strong AB oscillations are seen in the negative *excess cross-correlation* (the part of the cross-correlation above the background, resulting from an injected current of 0.3nA at each source). Note that the mean non-oscillating part of the excess cross-correlation is -1.2x10$^{29}$ A$^2$/Hz; as expected. Left panel: Similar averaging of the data but at a fixed $V_{MG}$. The somewhat different visibilities in both panels are due to analysis that must be done in different region of the 2D plot. (b) 2D FFT of the cross-correlation signal. A strong peak is visible with an integrated power 3.0×10$^{-30}$ A$^2$/Hz. (c) A vector representation of the different periodicities. The two vectors starting from the origin are the 2D periodicities of the two MZIs. The green cross is the 2D periodicity of the cross-correlation signal of the two-particle interferometer. The vectorial sum of the periodicities of the two MZIs (black dot) agrees excellently with the corresponding 2D periodicity of the two-particle interferometer.

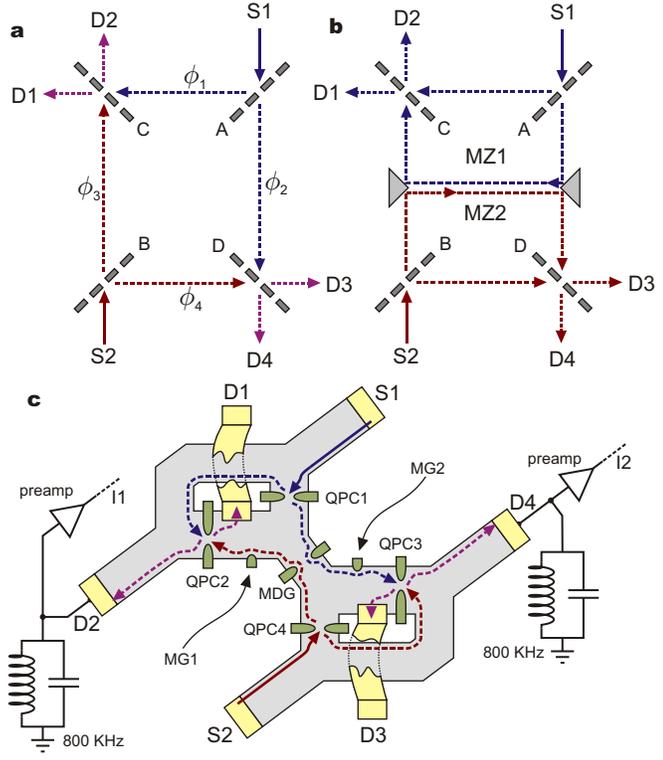
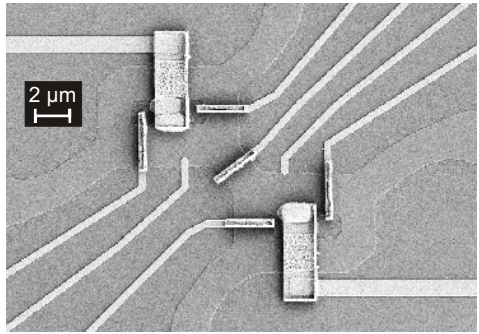

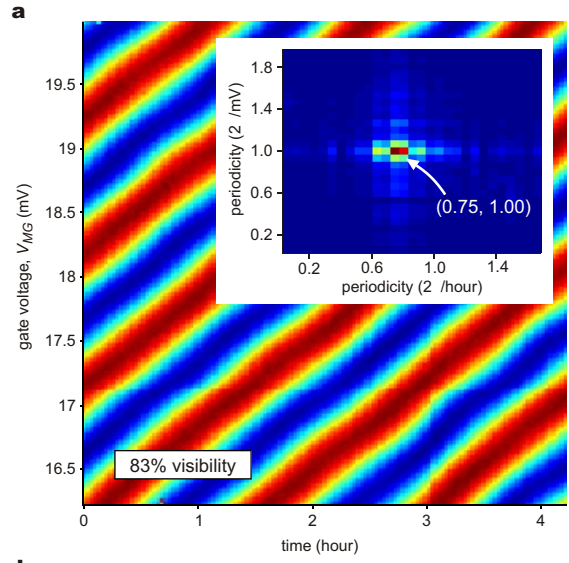

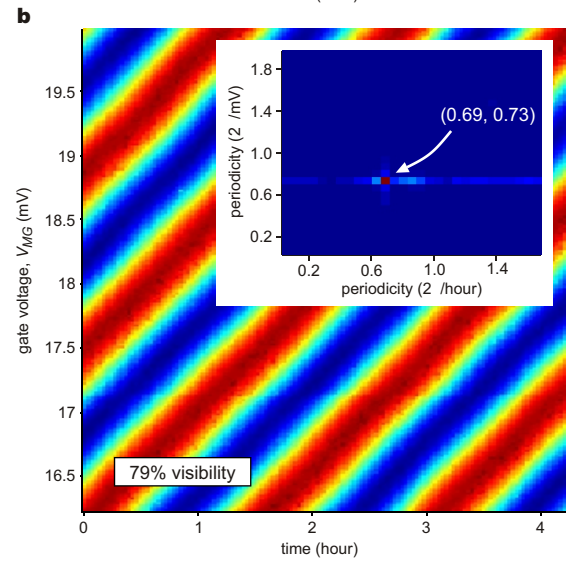

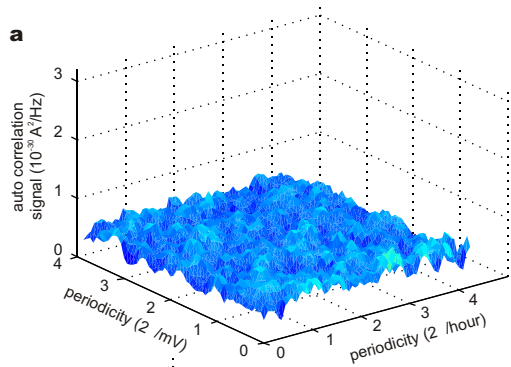

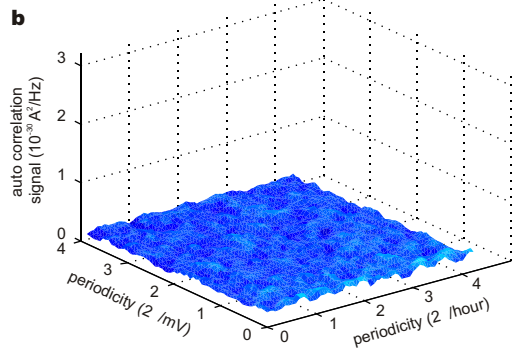

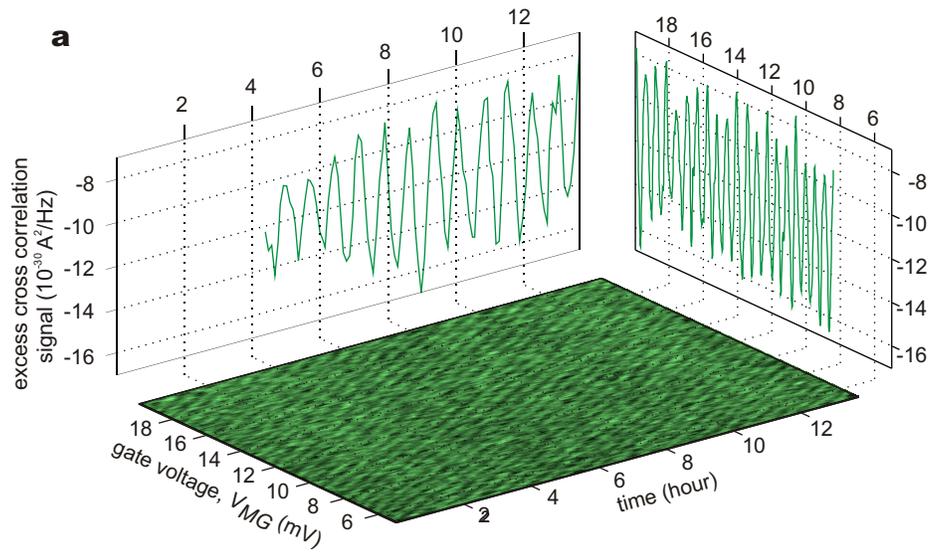
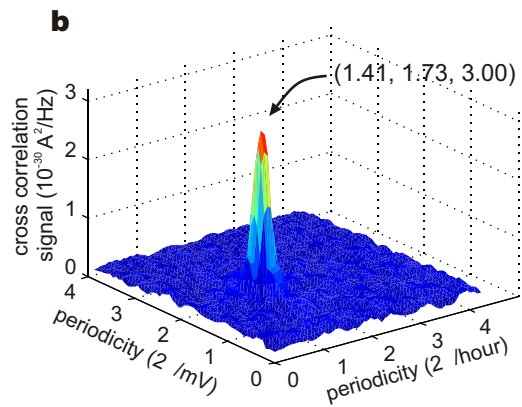
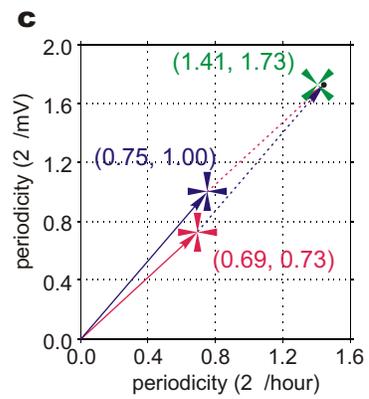